\documentclass[twocolumn,showpacs,aps,prl,superscriptaddress]{revtex4}

\usepackage{graphicx}
\usepackage{dcolumn}
\usepackage{amsmath}
\usepackage{epsfig}

\input pubboard/babarsym
\def\Dp      {\ensuremath{D^+}}
\def\Dm      {\ensuremath{D^-}}
\def\Dstar   {\ensuremath{D^{*}}}
\def\Dstarp  {\ensuremath{D^{*+}}}
\def\Dstarm  {\ensuremath{D^{*-}}}

\newcommand{\etal}{{\it et al.}}

\def\bsg    {\ensuremath {b \to s \gamma}}

\def\bkg    {\ensuremath {\B \to \Kstar \gamma}}
\def\bkog    {\ensuremath {\Bz \to \Kstarz \gamma}}

\def\bkpg    {\ensuremath {\Bp \to \Kstarp \gamma}}
\def\kpi    {\ensuremath {\Kstarz \to \Kp \pim}}

\def\Kz    {\ensuremath{K^{0}}\xspace}

\def\tg     {\ensuremath {\theta^{*}_T}}
\def\ctg     {\ensuremath {\cos{\tg}}}

\def\ebeam     {\ensuremath {E^{*}_{beam}}}
\def\esqbeam     {\ensuremath {E^{*2}_{beam}}}
\def\egcms     {\ensuremath {E^{*}_{\gamma}}}
\def\de        {\ensuremath {\Delta E^{*}}}
\def\mkpi      {\ensuremath {M_{\Kp \pim}}}
\def\mpl #1 #2 #3 {Mod.~Phys.~Lett.~{\bf#1},\ #2 (#3)}
\def\npb  #1 #2 #3 {Nucl.~Phys.~B~{\bf#1},\ #2 (#3)}
\def\plb  #1 #2 #3 {Phys.~Lett.~B~{\bf#1},\ #2 (#3)}
\def\pr   #1 #2 #3 {Phys.~Rep.~{\bf#1},\ #2 (#3)}
\def\prd  #1 #2 #3 {Phys.~Rev.~D~{\bf#1},\ #2 (#3)}
\def\prl  #1 #2 #3 {Phys.~Rev.~Lett.~{\bf#1},\ #2 (#3)}
\def\RMP  #1 #2 #3 {Rev.~Mod.~Phys.~{\bf#1},\ #2 (#3)}
\def\zpc  #1 #2 #3 {Z.~Phys.~C~{\bf#1},\ #2 (#3)}
\def\nim  #1 #2 #3 {Nucl.~Instrum.~Methods~{\bf#1},\ #2 (#3)}
\def\nima  #1 #2 #3 {Nucl.~Instrum.~Methods~A.{\bf#1},\ #2 (#3)}
\def\epjc #1 #2 #3 {Euro.~Phys.~Jour~{\bf#1},\ #2 (#3)}
\def\rmp #1 #2 #3 {Rev.~Mod.~Phys~{\bf#1},\ #2 (#3)}
\def\npbps #1 #2 #3 {Nucl.~Phys.~B.~proc.~suppl~{\bf#1},\ #2 (#3)}
\def\progtp #1 #2 #3 {Prog.~Theo.~Phys~{\bf#1},\ #2 (#3)}
\def\etal{{\it et al.}}

\def\brcharg {\ensuremath {\BR(\bkpg) = (3.83 \pm 0.62({\rm stat. }) \pm 0.22({ \rm sys. })) \times 10^{-5}}}
\def\brneut {\ensuremath {\BR(\bkog)= (4.23 \pm 0.40({\rm stat. }) \pm 0.22({ \rm sys. })) \times 10^{-5}}}

\def\aall {\ensuremath {A_{\CP}(\bkg) = -0.044 \pm 0.076({\rm stat. }) \pm 0.012({ \rm sys. }) }}
\def\aexcl {\ensuremath {  -0.170 < A_{\CP}(\bkg) <  0.082 }}

\def\lumi    {\ensuremath {(22.74 \pm 0.36 ) \times 10^6}}

\newcommand{\BABARPubYear}    {01}
\newcommand{\BABARPubNumber}  {04}

\newcommand{\SLACPubNumber} {8952}

\def\figurebox#1#2#3{%
    \def\arg{#3}%
    \ifx\arg\empty
    {\hfill\vbox{\hsize#2\hrule\hbox to #2{\vrule\hfill\vbox to #1{\hsize#2\vfill}\vrule}\hrule}\hfill}%
    \else
    {\hfill\epsfbox{#3}\hfill}%
    \fi}

\long\def\inst#1{\par\nobreak\kern 4pt\nobreak
    {\it #1}\par\vskip 10pt plus 3pt minus 3pt}

\begin{document}

\preprint{\babar-PUB-\BABARPubYear/\BABARPubNumber} 
\preprint{SLAC-PUB-\SLACPubNumber} 

\begin{flushleft}
\babar-PUB-\BABARPubYear/\BABARPubNumber\\
SLAC-PUB-\SLACPubNumber\\[10mm]
\end{flushleft}

\title{
{\large \bf
Measurement of  \bkg\  Branching Fractions 
and Charge Asymmetries} 
\begin{center} 
\vskip 10mm
The \babar\ Collaboration
\end{center}
}

%
\author{B.~Aubert}
\author{D.~Boutigny}
\author{J.-M.~Gaillard}
\author{A.~Hicheur}
\author{Y.~Karyotakis}
\author{J.~P.~Lees}
\author{P.~Robbe}
\author{V.~Tisserand}
\affiliation{Laboratoire de Physique des Particules, F-74941 Annecy-le-Vieux, France }
\author{A.~Palano}
\affiliation{Universit\`a di Bari, Dipartimento di Fisica and INFN, I-70126 Bari, Italy }
\author{G.~P.~Chen}
\author{J.~C.~Chen}
\author{N.~D.~Qi}
\author{G.~Rong}
\author{P.~Wang}
\author{Y.~S.~Zhu}
\affiliation{Institute of High Energy Physics, Beijing 100039, China }
% ...
\collaboration{The \babar\ Collaboration}
\noaffiliation

\date{October 25, 2001}

\begin{abstract}
The branching fractions of the  exclusive decays $\bkog$ and $\bkpg$ are measured
from a sample  of \lumi\ $\BB$ decays collected with the \babar\ detector
at the PEP II asymmetric $e^{+}e^{-}$ collider. We find $\brneut$,
$\brcharg$ and
constrain the $\CP$-violating charge asymmetry to be $\aexcl$ at 90\% C.L.
\end{abstract}


\maketitle

In the Standard Model (SM) the exclusive decays \mbox{\bkg} proceed dominantly by the electromagnetic  loop ``penguin'' transition $\bsg$. Many extensions of the SM provide new virtual high-mass fermions and bosons that can appear in the loop, causing deviations in the inclusive rate for $\bsg$~\cite{hewett}. The sensitivity of the exclusive rates to these effects is limited by the uncertainty in the SM calculation. However, there has been considerable progress recently~\cite{SM}. The precision measurement of
the exclusive branching fractions $\BR(\bkog)$, $\BR(\bkpg)$ is needed to test
and improve these calculations. The non-SM processes can also interfere with the SM decay to cause $\CP$-violating charge asymmetries at a level as high as 20\%~\cite{kagan}.
The $\CP$-violating charge asymmetry from SM contributions alone is expected to be $< 1\%$.

In this letter, measurements of the
exclusive branching fractions, $\BR(\bkog)$ in the $\Kstarz \to \Kp \pim$, $\KS\piz$ modes, and 
$\BR(\bkpg)$ in the  $\Kstarp \to \Kp \piz$, $\KS \pip$ modes
with $\KS \to \pip \pim$, are presented. Here $\Kstar$ refers to the $\Kstar(892)$ resonance and 
the charge conjugate decays are implied unless otherwise stated. The
$\Kstarz \to \Kp \pim$ and $\Kstarp \to \Kp \piz$, $\KS \pip$
modes are used to  search for $\CP$-violating charge asymmetries.

The data were collected with the \babar\ detector
~\cite{detector} at the \pep2\ asymmetric $\ep (3.1\gev ) $ -- $\en (9\gev ) $  storage ring~\cite{pep}. 
The results  in this paper are based upon an
integrated luminosity of 20.7\invfb\ of  data corresponding to
\lumi\ \BB\ meson pairs recorded at the $\FourS$ resonance (``on-resonance'') and
$2.6\invfb$ at $40 \mev$ below this energy (``off-resonance''). 

We use Monte Carlo simulations of the \babar\ detector 
based on GEANT 3.21~\cite{geant}  to optimize our selection criteria and to determine signal efficiencies.  Events taken from random triggers are used to measure the beam 
backgrounds. These simulations take account of varying detector conditions
and beam backgrounds.

The selection criteria for this analysis are optimized to maximize
$S^{2}/(S+B)$ where $S$ is the number of signal candidates expected, assuming
the central values of the previous measurement $\BR(\bkog,\,\bkpg) 
= (4.55^{+0.72}_{-0.68}(stat.)\pm 0.34 (sys.),3.76^{+0.89}_{-0.83}(stat.)\pm 0.28(sys.)) \times 10^{-5}$~\cite{cleo1}, and $B$ is the expected number
of background candidates determined from Monte Carlo and confirmed with
off-resonance data. Quantities are computed 
in both the laboratory frame and the center-of-mass  frame of the $\epem$ system. Those 
computed in the center-of-mass frame are denoted by an asterisk; e.g. $\ebeam=5.29\gev$ is the
on-resonance energy of the \ep\ and \en\ beams.

We require a high-energy radiative photon candidate with 
energy  $1.5 < E_{\gamma} < 4.5\gev $ in the laboratory frame  and $2.30  < \egcms < 2.85\gev $ in the center-of-mass  frame.
A photon candidate is defined as a localized energy 
maximum~\cite{detector} in the
calorimeter acceptance  $-0.74 <\cos{\theta} < 0.93$, where $\theta$
is the polar angle to the detector axis. It must be 
isolated by 25 cm from any other photon candidate or track and have a  lateral energy profile consistent with a photon shower. We veto photons from a $\piz(\eta)$ by requiring that
the invariant mass of the combination with any other photon of energy greater than 
50 (250)\mev\ not be within the range 
$115(508) < M_{\gamma\gamma} < 155(588) \mevcc$.

The $\Kstar$ is reconstructed from $\Kp$, $\KS$, $\pim$ and $\piz$ candidates through the 
four modes $\Kstarz\to\Kp\pim,~\KS\piz$ and $\Kstarp\to\Kp\piz,~\KS\pip$.
The  $\Kp$ and $\pim$ track candidates  are required to be well reconstructed in
the drift chamber and to originate from a vertex consistent with the
$\epem$ interaction point (IP). The $\KS$ candidates are reconstructed from two
oppositely-charged tracks coming from a common vertex displaced from
the IP by at least 0.2\cm\ in the transverse plane  and having an invariant mass 
$489 < M_{\pip \pim} < 507\mevcc $. A track is identified as a kaon if it
is projected to pass through the fiducial volume of the particle identification
detector, an internally-reflecting ring-imaging Cherenkov detector (DIRC)~\cite{detector}, and the cone of Cherenkov light is consistent 
in time and angle with a kaon of the measured track momentum.
A charged pion is identified as a track that is not a kaon.
  The \piz\ candidates are reconstructed from pairs of photons, each
with energy greater than $30\mev$, and are  required to have $115 < M_{\gamma \gamma} < 150 \mevcc$ and $E_{\piz} > 200 \mev$. A mass-constraint fit to the nominal \piz\ mass is used to improve the resolution of its momentum.
 The $\Kstar$ reconstruction is completed by requiring
the invariant mass of the candidate pairs  to be within 100\mevcc\ of the $\Kstarz$/$\Kstarp$ mass.

\begin{figure}[!htb]
\begin{center}
\includegraphics[width=0.9\linewidth]{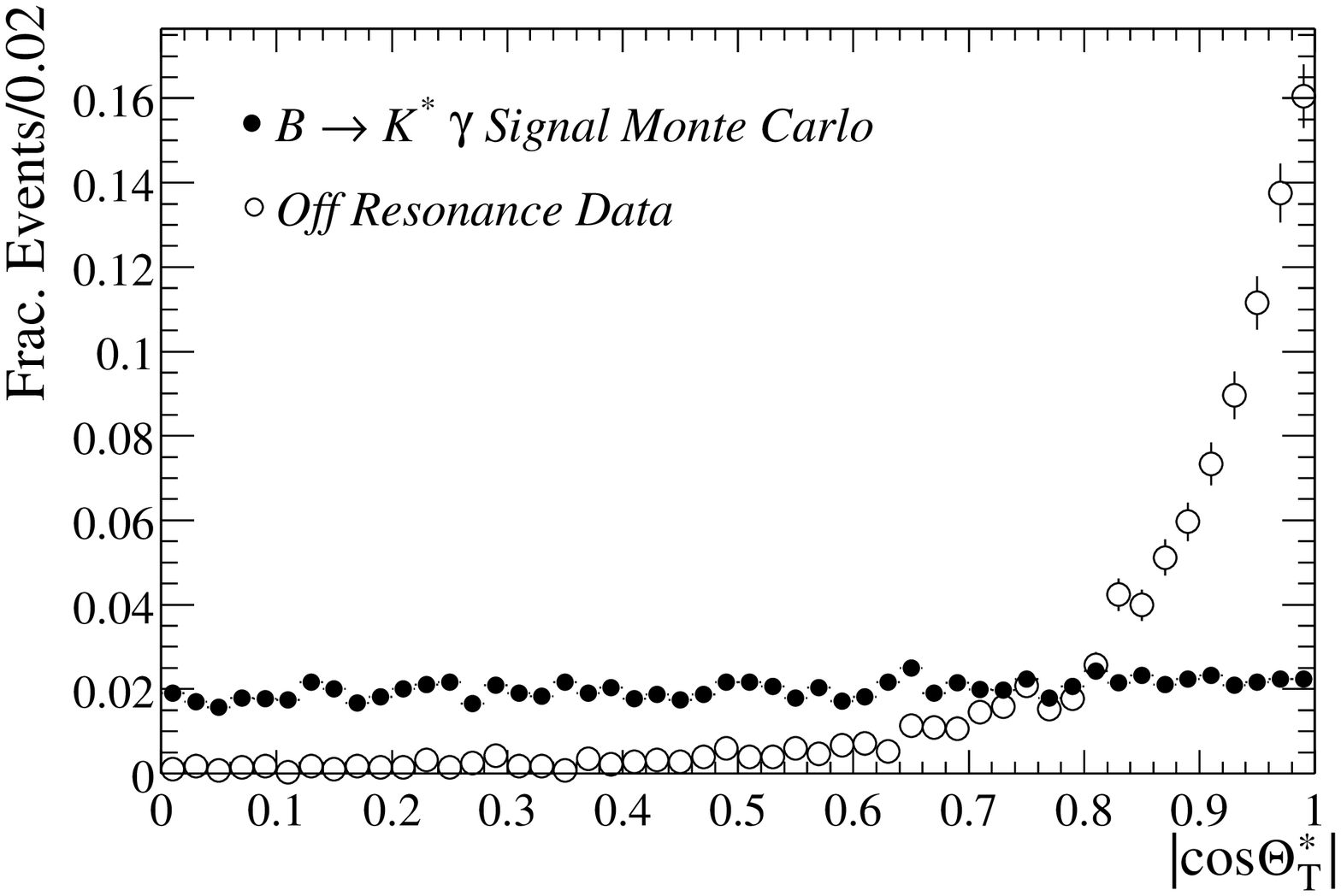}
\caption{The event shape variable $|\ctg|$ for $\bkog$, \kpi\ Monte Carlo and
off-resonance data.}
\label{fig:costhr}
\end{center}
\end{figure}

The $B$ meson candidates are reconstructed from the \Kstar\ and $\gamma$
candidates. The background is predominantly 
from continuum \qqbar\ production, with the
high-energy photon originating  from  initial-state radiation or from 
$\piz$ and $\eta$ decays. The background from other $B$ meson decays is
found to be negligible from Monte Carlo simulation. We exploit event topology differences between signal
and background to reduce the continuum contribution. We compute the thrust 
axis of the event  excluding the $B$ meson daughter candidates. 
Figure~\ref{fig:costhr} shows the distribution of $|\ctg|$ for
signal Monte Carlo events and off-resonance data, where \tg\ is the angle 
between the high-energy photon candidate and 
the thrust axis.
In the center-of-mass frame, \BB\ pairs are
produced approximately at rest and  produce a uniform $|\ctg|$ distribution.
In contrast,
\qqbar\ pairs are produced back-to-back in the center-of-mass frame which results in a $|\ctg|$ distribution peaking at
$1$. 
We require $|\ctg| < 0.8$. We further suppress backgrounds using the
angle of the $B$ meson candidate's direction with respect to the 
beam axis, $\theta^{*}_{B}$,  and the helicity angle of the \Kstar\  decay, $\theta_{H}$. The helicity angle is defined as the angle between either one of the \Kstar\ daughters' momentum vectors computed in the rest frame of the \Kstar\ and the \Kstar\ momentum vector in the parent $B$ meson rest frame. It follows a $\sin^{2}\theta_{H}$ distribution for the signal and peaks slightly towards $\pm 1$ for \qqbar\ background. The $B$ meson candidate's direction also follows a $\sin^{2}\theta^{*}_{B}$ for
the signal and is approximately flat for the \qqbar\ background. We require  
$|\cos\theta^{*}_{B}| < 0.80$ and $|\cos\theta_{H}| < 0.75$. 

Since the $B$ mesons are produced via $\epem \to \Y4S \to \BB$, the energy of
the $B$ meson in the center-of-mass frame is the beam energy, {$\ebeam$}. This is compared
to the measured energy of the $B$ meson daughters by defining 
$\de = E^{*}_{K^*} + \egcms - \ebeam $. The distribution of \de\ is
peaked at zero for the signal with a width dominated by the resolution of the photon candidates. It is asymmetric due to energy leakage from the calorimeter. 
We require $ -200  < \de < 100\mev $ 
for the  $\Kp \pim$, $\KS \pip$ modes and   $ -225  < \de < 125\mev $ for 
the modes containing a $\piz$, namely $\Kp \piz$ and $\KS\piz$.
The beam-energy substituted mass is  defined as 
$\mes= \sqrt{ E^{*2}_{beam}-\mbox{\boldmath $\mathrm p$}_{B}^{*2}}$, where \mbox{\boldmath $\mathrm p$}$^{*}_{B}$ is the momentum vector of the $B$ meson candidate calculated from the measured momenta of the daughters. The $\mes$ distribution for the signal is well
described by an asymmetric resolution function~\cite{novo}, with an
approximately Gaussian core  dominated by the resolution of
the beam energy measurement, and an asymmetric tail caused by the energy leakage from the calorimeter for the photon candidates. For the modes 
containing a single photon candidate, namely $\Kp \pim$ and $\KS \pip$,
we can remove the tail in $\mes$ by rescaling the measured photon energy \egcms\ by a factor $\kappa$, determined for each event, so that 
$E^{*}_{K^*}+\kappa\egcms - E_{\rm beam}^{*}=0$. The signal for these modes is then
described by a Gaussian. The background is empirically 
described by a threshold function~\cite{argusf} for each mode. We select candidates with $\mes >5.2\gevcc$.

\begin{figure}[!htb]
\begin{center}
\includegraphics[width=0.9\linewidth]{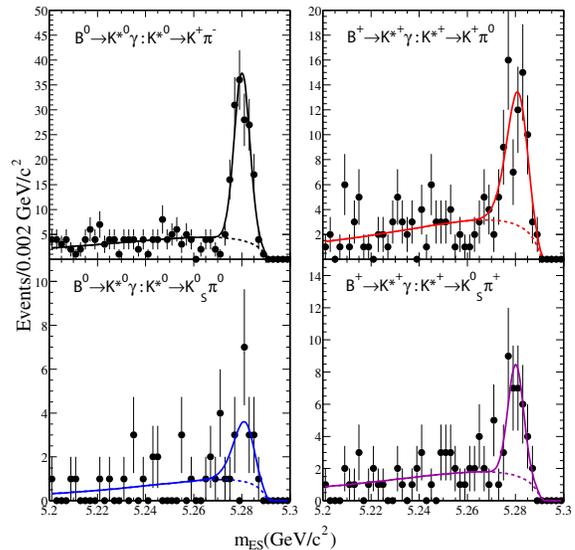}
\caption{\mes\ for the \bkg\ candidates. The fits used to extract
the signal yield are described in the text.}
\label{fig:datamb}
\end{center}
\end{figure}

Figure~\ref{fig:datamb} shows
the \mes\ distribution for each of the four modes. 
An unbinned maximum-likelihood technique is used
to fit the \mes\ distributions
for signal~\cite{novo} and background~\cite{argusf} contributions. The signal mean and width are allowed to vary in the fit for the high statistics $\Kp\pim$ and $\Kp\piz$ modes. 
The fitted width is slightly larger than the predicted 
Monte Carlo value. 
In the lower statistics modes we fix the width 
to the Monte Carlo value adjusted for the small
difference observed in the high statistics modes.
We fit the on-resonance data with a signal plus background shape, and 
simultaneously the on-resonance sideband and off-resonance samples with
the same background function, using a common fit parameter.
The off-resonance data sample is required to pass the same selection
criteria as the on-resonance data sample except that we remove the kaon
particle identification requirement to gain statistics in the $\Kp\pim$ and $\Kp\piz$ modes. The
on-resonance sideband sample is selected with the same criteria as the on-resonance data sample, 
except that we require $150 < \de < 400 \mev$ in the $\KS\piz$ and $\Kp\piz$ modes, and
$100 < \de < 500 \mev$ in the $\Kp\pim$ and $\KS\pip$ modes.  The signal yields with statistical errors from the fit are given in Table~\ref{tab:fit}.

\begin{table}[!htb]
\caption{The fitted signal yield, efficiency (including $\BR(\Kstar)$ and $\BR(\Kz)$),
cross-feed and down-feed from other penguin decays and measured 
branching fraction $\BR(\bkg)$ for each of the decay modes.  }
\begin{center}
\mbox{ \scriptsize
\begin{tabular}{lrcccl} \hline \hline
Mode        & Effici     &  \#Signal           & \#Cross  & \#Down   & $\BR(\bkg)$ \\
            & -ency      &  events             & -feed  &-feed   & $\pm {\rm stat.} \pm {\rm { \rm sys. }}$    \\ 
            &   \%       &                     & events & events & $\times 10^{-5}$    \\ \hline
$\Kp\pim$   &  $14.0$    & \hspace{0.1em}$135.7 \pm$ 13.3  & $0.4 \pm 0.1$    & $0.6 \pm 0.1$  & $4.24 \pm 0.41 \pm 0.22$   \\ 
$\KS\piz$   &  $1.4$     &  $14.8\pm  5.6$  & $0.4 \pm 0.1$    & $1.0 \pm 0.2$  & $4.10 \pm 1.71 \pm 0.42$   \\
$\KS\pip$   &  $3.9$     &  $28.1 \pm  6.6$  & $0.7 \pm 0.2$   & $1.2 \pm 0.2$  & $3.01 \pm 0.76 \pm 0.21$   \\
$\Kp\piz$   &  $4.3$     &\hspace{0.2em}  $57.6 \pm 10.4$  & $1.2 \pm 0.2$    & $2.6 \pm 0.4$   & $5.52 \pm 1.07 \pm 0.38$   \\ \hline \hline
\end{tabular}}
\end{center}
\label{tab:fit}
\end{table}

\begin{figure}[!htb]
\centerline{
\includegraphics[height=4.5cm,width=4.3cm]{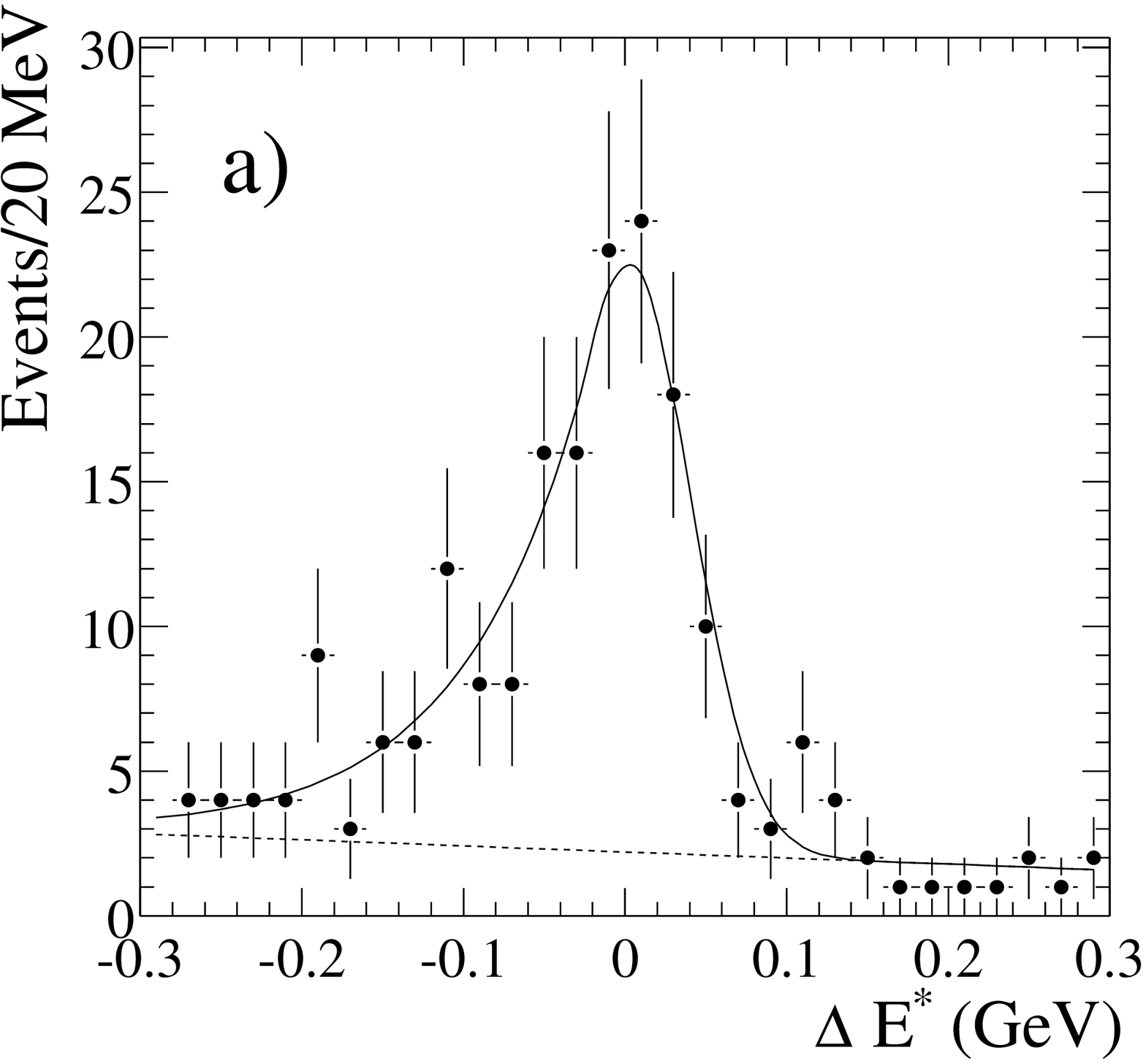}
\includegraphics[height=4.5cm,width=4.3cm]{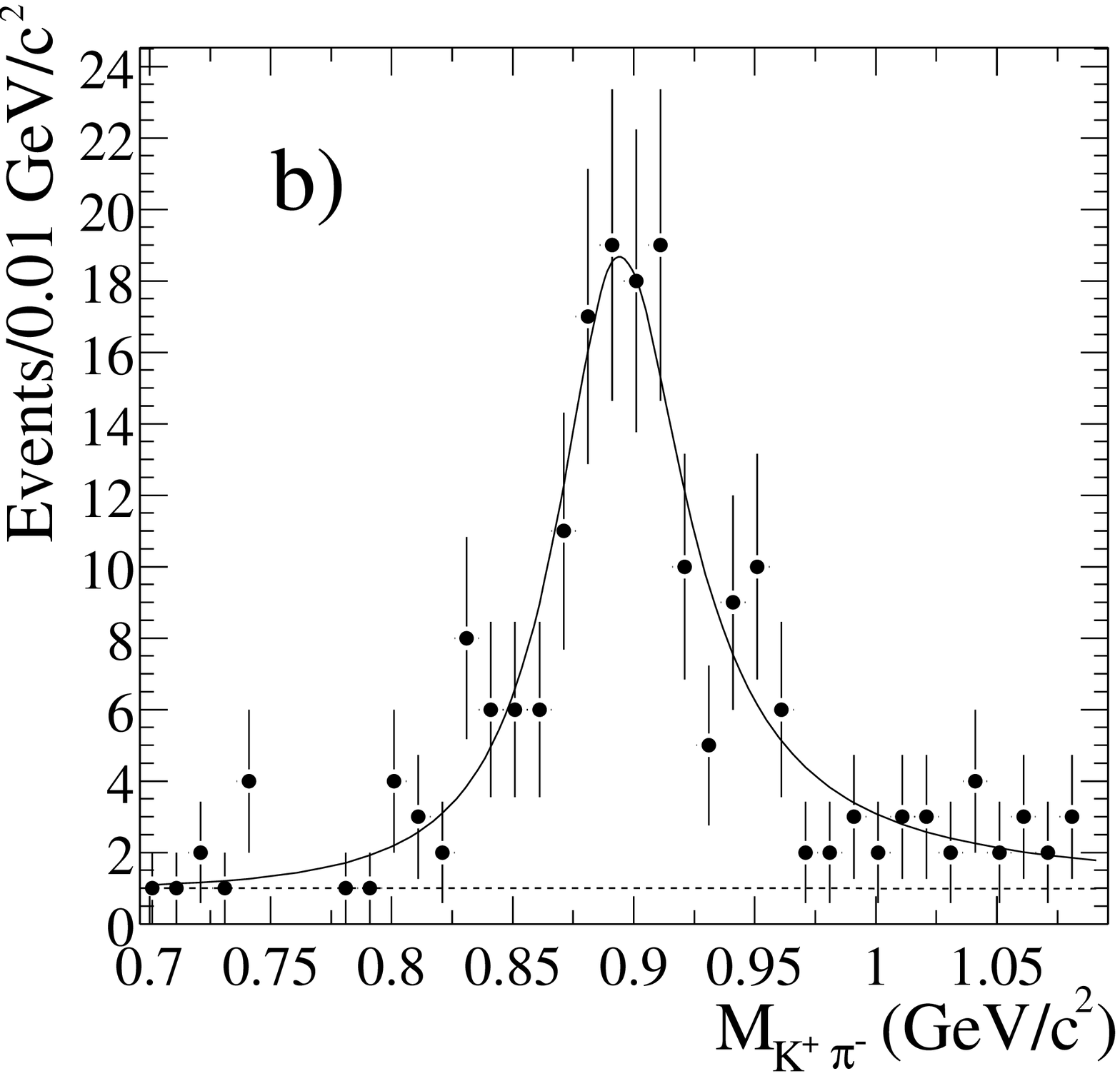}}
\caption{a) The \de\ projection for $\bkog$, $\Kstarz \to \Kp \pim$ candidates. The curve is the Monte Carlo expectation with a linear background.  b)  The \mkpi\ projection for $\bkog$, $\Kstarz \to \Kp \pim$ candidates with the \mkpi\ mass cut relaxed. The curve
is a fit to a relativistic Breit-Wigner function with linear background.}
\label{fig:datade}
\end{figure}

As a consistency check we plot in Figure~\ref{fig:datade}a the \de\ projection
for the $\Kp \pim$ mode after requiring $ 5.27  < \mes < 5.29\gevcc $.  
A comparison of the observed \de\ distribution with Monte Carlo shows 
good agreement.
We also plot \mkpi\ in Figure~\ref{fig:datade}b after
requiring $ 5.27  < \mes < 5.29\gevcc $, 
$ -200  < \de < 100\mev $ and $0.7 < \mkpi < 1.1\gevcc$. We fit with a relativistic Breit-Wigner plus
linear background shape and determine that the signal is consistent with coming from the $\Kstar(892)$. 

The efficiency for the selection of $\bkg$ candidates is given in Table~\ref{tab:fit}. The branching fraction is determined from the yield, the efficiency
and the total number of \BB\ events in the sample. The
cross-feed from the other $\bkg$ modes  and the down-feed from $B \to X_{s} \gamma$ are
estimated with Monte Carlo assuming the measured branching fractions from the
CLEO collaboration~\cite{cleo1, cleo2} for each mode and subtracted from the signal yield. 

The total systematic error  is the sum in quadrature  of the 
components shown in Table~\ref{tab:systematic}. The
systematic uncertainty in the signal yield derives from
uncertainties in the signal line shape, and cross-feed and down-feed contributions. 
The uncertainty in the signal line shape results from the $m_{ES}$ width difference
described above.
To gain statistics in the off-resonance data sample
used to fit the background function for the  $\Kp\pim$ and $\Kp\piz$ modes we
relax the kaon identification requirement and consequently 
assign a systematic uncertainty to the assumption that the background shape
is unaffected. 
The error in the assumed branching fractions and final-state modeling for
$B \to X_{s}\gamma$~\cite{cleo2} gives a systematic error in the
estimated down-feed from these modes.
The tracking efficiency is computed by identifying tracks in the silicon
vertex detector and observing the fraction that is well reconstructed in the drift chamber. 
We estimate the \KS\ efficiency uncertainty
by comparing the momenta and flight-distance distributions in data and
Monte Carlo. The kaon identification efficiency in the DIRC is derived from a sample of 
$D^{*+} \to D^{0}\pip,D^{0} \to \Km\pip$ decays. 
The photon and $\piz$ efficiencies are measured
by comparing the ratio of events 
$N(\tau^{\pm} \to h^{\pm} \piz)/N( \tau^{\pm} \to h^{\pm} \piz \piz)$ 
to the previously measured branching ratios \cite{cleo3}. 
The photon isolation and $\piz/\eta$ veto efficiency
are dependent on the event multiplicity and are tested by ``embedding'' 
Monte Carlo-generated photons into both an exclusively reconstructed $B$ meson data sample  and
a generic $B$ meson Monte Carlo sample. The \de\ resolution is dominated by the photon-energy resolution so that uncertainties in the calorimeter energy resolution and overall energy-scale cause
an uncertainty in the efficiency of the \de\ requirement. The photon-energy resolution is measured in data using $\piz$ and $\eta$ meson decays and  
$\epem \to \epem \gamma$ events. The energy scale uncertainty 
is estimated by using a sample of $\eta$ meson decays with 
 symmetric energy photons; the deviation in the reconstructed $\eta$ 
mass from the nominal $\eta$ mass provides an estimate of the  uncertainty in the measured single photon energy.

\begin{table}[!htb]
\caption{The systematic uncertainties in the measurement of
\BR(\bkg).}
\begin{center}
\begin{tabular}{lcccc}  \hline \hline
                  & \multicolumn{4}{c}{\% Uncertainty in } \\ 
                  & \multicolumn{4}{c}{   \BR(\bkg) }  \\ [2pt] \cline{2-5} 
                                        & $\Kp\pim$ & $\KS \piz$       & $\KS \pip$ & $\Kp \piz$ \\  \hline

\mes\ Line shape                        &   -       & 7.4              & 1.7     & 1.9   \\
Background shape                        &  1.0      &  -               &  -      & 3.8         \\
Down-feed modeling                     &  1.0      & 1.5              & 1.0     & 1.2          \\
$K^{\pm}/\pi^{\pm}$ tracking efficiency &  2.4      &  -               & 1.2     & 1.3          \\
\KS\ efficiency                         &   -       & 4.5              & 4.5     &  -         \\
Kaon identification                     &  0.7      &  -               &  -      & 1.0          \\
Photon efficiency                       &  1.3      & 1.3              & 1.3     & 1.3       \\
Photon distance cut                     &  2.0      & 2.0              & 2.0     & 2.0        \\
\piz\ efficiency                        &   -       & 2.5              &  -      & 2.5          \\
$\piz$/$\eta$ veto                      &  1.0      & 1.0              & 1.0     & 1.0       \\
Energy resolution                       &  2.5      & 2.5              & 2.5     & 2.5       \\
Energy scale                            &  1.0      & 1.0              & 1.0     & 1.0       \\
MC statistics                           &  1.9      & 2.4              & 1.5     & 2.1          \\
$B$ counting                            &  1.6      & 1.6              & 1.6     & 1.6           \\ \hline
Total                                   &  5.3      & 10.3             & 6.7     & 7.0    \\ \hline \hline
\end{tabular}
\end{center}
\label{tab:systematic}
\end{table}

   The \bkg\ samples, except for the $\KS\piz$ sample,  are used to search for $\CP$-violating charge
asymmetries $A_{\CP}$, defined by
\[
A_{\CP}= \frac{\Gamma(\Bbar \to \Kstarb \gamma) - \Gamma(\B \to \Kstar\gamma) }{\Gamma(\Bbar \to \Kstarb \gamma) + \Gamma(\B \to \Kstar \gamma)  } \,.
\]
The flavor of the underlying $b$ quark is tagged  by the charge
of the $\Kpm$ or $\Kstarpm$ in the decay. The on-resonance sample for
each mode is divided into two $\CP$-conjugate samples and the
signal yield for each is extracted with the same fitting technique 
as for the branching fraction measurements. In the fit the
background shape and normalization, as well as the signal peak and width are constrained to be the same for both $\CP$-conjugate samples. The measured asymmetries and  the asymmetry of the background in the
sideband regions defined by $ -200  < \de < 100 \mev$, $5.2   < \mes < 5.27 \gevcc $ are given in Table~\ref{tab:cpv}.
\begin{table}[!htb]
\caption{ The measured $A_{\CP}$ in signal and background samples.}
\begin{center}
\begin{tabular}{lcc} \hline \hline
Mode        & $A_{\CP}({\rm signal})$      &  $A_{\CP}({ \rm background })$   \\  
            & ($\pm {\rm stat. } \pm { \rm sys. }$)&  ($\pm {\rm stat. }$) \\ \hline  
 $\Kp\pim$  & $-0.049 \pm 0.094 \pm 0.012$      &  $-0.011 \pm  0.104$   \\ \hline 
$\KS\pip$   & $-0.190 \pm 0.210 \pm 0.012$        &  $-0.080  \pm 0.080$      \\
 $\Kp\piz$  & \hspace{0.8em}$ 0.044 \pm 0.155 \pm 0.021$       &  $\hspace{0.0em}-0.022 \pm 0.105 $   \\ \hline \hline
\end{tabular}
\end{center}
\label{tab:cpv}
\end{table}

The systematic uncertainty in the asymmetry is due to possible detector
effects that cause a different reconstruction efficiency for the
two \CP\-conjugate decays. This uncertainty has been estimated in data
with a  number of known charge-symmetric processes to be less than $ 2 \% $.


Finally, we combine the measured branching fractions for the individual
modes using a weighted average, $\brcharg$, $\brneut$. The weighted average of the measured $\CP$-violating charge asymmetries is $\aall$. We
constrain $\aexcl$ at $90\%$ C.L.

We are grateful for the excellent luminosity and machine conditions
provided by our \pep2\ colleagues.
The collaborating institutions wish to thank 
SLAC for its support and kind hospitality. 
This work is supported by
DOE
and NSF (USA),
NSERC (Canada),
IHEP (China),
CEA and
CNRS-IN2P3
(France),
BMBF
(Germany),
INFN (Italy),
NFR (Norway),
MIST (Russia), and
PPARC (United Kingdom). 
Individuals have received support from the Swiss NSF, 
A.~P.~Sloan Foundation, 
Research Corporation,
and Alexander von Humboldt Foundation.


\begin{thebibliography}{99}
\bibitem{novo}
The signal for the $\Kp\piz$ and $\KS\piz$ modes 
is fit with the function
$
dN/d\mes = A_S 
\exp{\left[-
  0.5(\ln^2(1+\Lambda\tau(\mes-m_0))/\tau^2
         +\tau^2)   
\right]}
$
where
$\Lambda=\sinh(\tau\sqrt{\ln{4}})
/(\sigma\tau\sqrt{\ln{4}})$,
the peak position is 
$m_0$, the width is $\sigma$, and
$\tau$ is the tail parameter. The $\Kp\pim$ and $\KS\pip$ modes
are fitted with a Gaussian distribution.
\end{thebibliography}
\end{document}